\documentclass[journal, a4, 11pt, twoside, twocolumn]{IEEEtran}

\usepackage{newtxmath}
\usepackage{amsmath}
\usepackage{amsfonts}
\usepackage{mathtools, nccmath}

\usepackage[superscript, noadjust]{cite}

\usepackage{graphicx}
\graphicspath{{figures/}}
\DeclareGraphicsExtensions{.pdf,.jpeg,.png,.eps}

\usepackage{booktabs}
\usepackage{array}
\usepackage{footnote}
\usepackage[acronym]{glossaries}
\usepackage{color}
\usepackage{enumitem}
\usepackage{stfloats}
\usepackage[hidelinks]{hyperref}
\usepackage[protrusion, expansion, kerning, spacing]{microtype}

\usepackage{siunitx}
\newacronym{snn}{SNN}{Spiking Neural Network}
\newacronym{dnn}{DNN}{Deep Neural Network}
\newacronym{stdp}{STDP}{Spike-Timing Dependent Plasticity}
\newacronym{bptt}{BPTT}{Back-Propagation Through Time}

\title{Neurobench: DCASE 2020 Acoustic Scene Classification benchmark on Xylo™Audio 2}
\author{Weijie Ke $^{1}$, Mina Khoei$^{1}$, Dylan Muir$^{1}$ %
    \thanks{$^1$SynSense}%
}
\date{v1.0 September 2024}

\begin{document}

\maketitle

\begin{abstract}
Xylo™Audio is a line of ultra-low-power audio inference chips, designed for in- and near-microphone analysis of audio in real-time energy-constrained scenarios.
Xylo is designed around a highly efficient integer-logic processor which simulates parameter- and activity-sparse spiking neural networks (SNNs) using a leaky integrate-and-fire (LIF) neuron model.
Neurons on Xylo are quantised integer devices operating in synchronous digital CMOS, with neuron and synapse state quantised to \SI{16}{\bit}, and weight parameters quantised to \SI{8}{\bit}.
Xylo is tailored for real-time streaming operation, as opposed to accelerated-time operation in the case of an inference accelerator.
Xylo™Audio includes a low-power audio encoding interface for direct connection to a microphone, designed for sparse encoding of incident audio for further processing by the inference core.

In this report we present the results of DCASE 2020 acoustic scene classification audio benchmark dataset deployed to Xylo™Audio~2.
We describe the benchmark dataset; the audio preprocessing approach; and the network architecture and training approach.
We present the performance of the trained model, and the results of power and latency measurements performed on the Xylo™Audio~2 development kit.

This benchmark is conducted as part of the Neurobench project\cite{yik2024neurobench}.
\end{abstract}


\section*{Benchmark dataset}

The benchmark is based on the DCASE 2020 acoustic scene classification challenge, utilizing the TAU Urban Acoustic Scenes 2020 Mobile datasets (development and evaluation)\cite{Kumari_2019_IPDPSW}.
It features recordings from 12 European cities across 10 different acoustic scenes, captured using four different devices.
Recordings were made using four devices that captured audio simultaneously.
The primary recording device, referred to as device A, consists of a Soundman OKM II Klassik/studio A3 electret binaural microphone paired with a Zoom F8 audio recorder, utilizing a 48kHz sampling rate and 24-bit resolution. The other devices are commonly available consumer electronics: device B is a Samsung Galaxy S7, device C is an iPhone SE, and device D is a GoPro Hero5 Session.
The acoustic scenes included in the dataset are: airport, Indoor shopping mall, Metro station, pedestrian street, public square, street with medium traffic, travelling by tram, travelling by bus, travelling by underground metro, urban park.

The cities where the audio data was recorded are Amsterdam, Barcelona, Helsinki, Lisbon, London, Lyon, Madrid, Milan, Prague, Paris, Stockholm, and Vienna.
The dataset was collected by Tampere University of Technology from May 2018 to November 2018. 
Out of the 10 available scene classes, 4 are used: ``airport,'' ``street with medium traffic,'' ``travelling by bus,'' and ``urban park.''
Among the 4 real and simulated recording devices, only one real device ``Device A'' is used.
The audio samples are divided into 1-second segments with no resampling requirements or limitations.
The dataset includes splits for training (\num{41360} samples), validation (\num{1320} samples), and evaluation (\num{16240} samples).

\section*{Audio preprocessing}
\begin{figure}
    \centering
    \includegraphics[width=\linewidth]{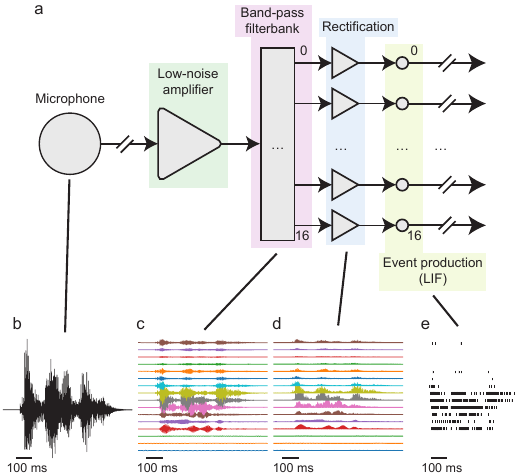}
    \caption{
        \textbf{Audio preprocessing approach\cite{bos2024micropower}.}
        \textbf{a} The stages of audio preprocessing in Xylo™Audio 2.
        Single-channel audio arrives at a microphone (\textbf{b}).
        This passes through a band-pass Butterworth filterbank, and is split into $N=16$ frequency bands (\textbf{c}).
        Filter output is rectified (\textbf{d}) before passing through a bank of LIF neurons that smooth and quantize the signals in each band.
        The result is a set of sparse event channels (\textbf{e}), where the firing intensity in each channel is proportional to the instantaneous energy in each frequency band.
    }
    \label{fig:audio_preprocessing}
\end{figure}

We encoded each sample as sparse events, using a simulation of the audio encoding hardware present on the Xylo™Audio device.
The design of this preprocessing block is shown in Figure~\ref{fig:audio_preprocessing}\cite{muir_rockpool_2019, bos2024micropower}.
Briefly, this block is a streaming-mode buffer-free encoder, designed to operate continuously on incoming audio.
A low-noise amplifier with a selectable gain of \num{0}, \num{6} or \SI{12}{\deci\bel} amplifies the incoming audio.
A band-pass filter bank with 2nd-order Butterworth filters splits the signal into 16 bands, with centre frequencies spanning \SIrange{40}{16940}{\hertz} and with a Q of 4.
The output of these filters is rectified, then passed through a leaky integrate-and-fire (LIF) neuron to smooth the signal and convert it to events.
The result is to convert a single audio channel into 16 sparse event channels with event rate in each channel corresponding to the energy in each frequency band.

Samples were using the preprocessing block described here, and binned temporally to \SI{10}{\milli\second}.
Our approach operates in streaming-mode, analysing a continuous time-frequency representation of the input audio, similar to a real-time Fourier transform (see Figure~\ref{fig:audio_preprocessing}e).

\section*{Network architecture}

\begin{figure*}
    \centering
    \includegraphics[width=\textwidth]{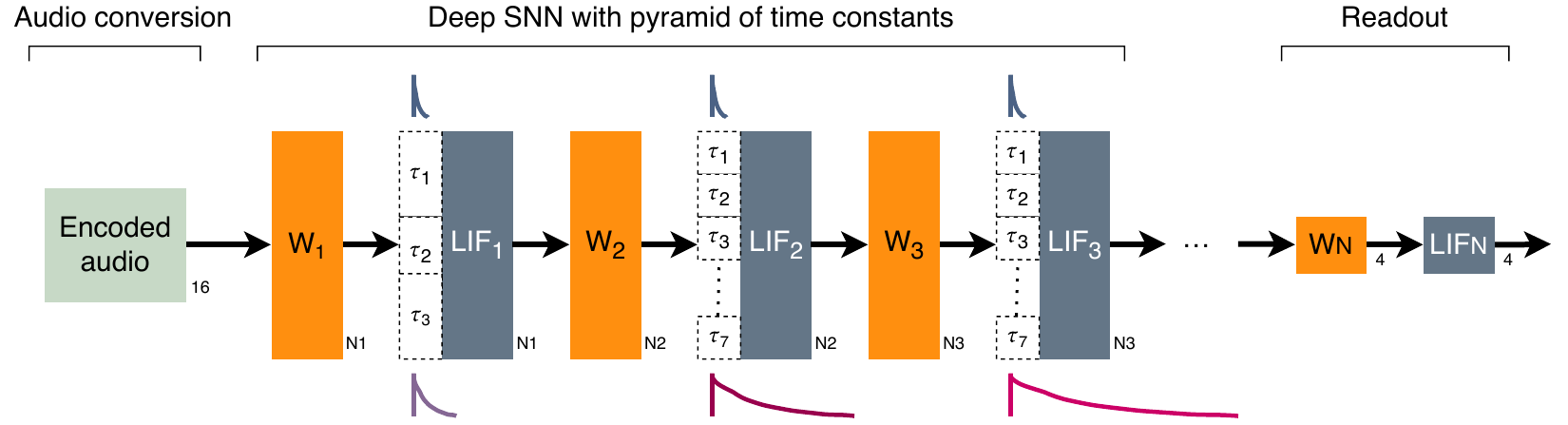}
    \caption{
        \textbf{The SynNet architecture used in this benchmark\cite{bos_sub-mw_2022, bos2024micropower}.}
        Event-encoded audio is provided as input, as described in Figure~\ref{fig:audio_preprocessing}.
        The network consists of a single feed-forward chain of fully-connected layers, using the LIF neuron model.
        Several time constants are distributed over each layer, with shorter time constants in early layers and longer time constants in later layers (see text for details).
        Four readout readout LIF neuron is used in each network.
    }
    \label{fig:synnet_architecture}
\end{figure*}

We use a feed-forward spiking neural network architecture called ``SynNet''\cite{bos_sub-mw_2022, bos2024micropower} (Figure~\ref{fig:synnet_architecture}).
This is a fully-connected multi-layer architecture, interleaving linear weight matrices with LIF neuron layers.
Each layer has a number of synaptic time constants, where the time constants are defined as $\tau_n = 2^n * \SI{10}{\milli\second}$, and neurons are distributed with the range of time constants for that layer.
For instance Layers with 3 time constants have one third of the neurons with synaptic time constants $\tau_1 = \SI{20}{\milli\second}$; one third with $\tau_2 = \SI{40}{\milli\second}$; and one third  with $\tau_3 = \SI{80}{\milli\second}$ and so on.
Membrane time constants for all neurons, as well as readout neuron time constants, are set as $\tau_m = \SI{20}{\milli\second}$.

For example, the network presented here used the parameters $H = \left[31, 31, 31 \right]$ $\tau = \left[3, 7, 7\right]$; 3 hidden layers with a first hidden layer width of 31 neurons, followed by 31 neurons, and so on; and with the first hidden layer containing 3 synaptic time constants, the second with 7 synaptic time constants, the third with 7 and so on.
Four readout LIF neurons was present, corresponding to the four target classes.
The network was trained such that the output neuron with highest firing rate indicated the detected class.

\section*{Training}

Networks were defined using the open-source Rockpool toolchain (\url{https://rockpool.ai} \cite{muir_rockpool_2019}), with the \textit{torch} back-end.
During training, the membrane potential of the readout neurons was taken as the network output.
The training loss for readout channels was defined as follows.
\begin{align*}
    \textrm{PeakLoss}(\mathbf{x}_i, y) & =
    \begin{cases}
        \textrm{MSE}\left(1/M\int_m^{m+M}\mathbf{x}_i, \textbf{g}\right) \hfill \textrm{if } i = y\\
        w_l \cdot \textrm{MSE}\left(\mathbf{x}_i, \textbf{0}\right) \hfill \textrm{otherwise}
    \end{cases}
\end{align*}
where
$\mathbf{x}_i$ is a membrane potential vector over time for each of readout channels $ i \in \{0,1,2,3\} $;
 As targets are one hot coded vectors for 4 classes, then:
$$ y = {\mathrm{argmax}}(\textit{target}) $$
MSE is the mean-squared-error loss function;
$m = {\mathrm{argmax}(\mathbf{x})}$, is the index of the peak value in $\mathbf{x}_i$.
$M$ is the window duration to examine from $\mathbf{x}_i$ following the peak;
$\textbf{g}$ is a vector $g\cdot \textbf{1}$, a target value that $\textbf{x}$ should match around its peak;
$w_l$ is a weighting for the non-target loss component;
$\textbf{0}$ is the vector of all-zeros.
For the networks trained here, we took $M=\SI{100}{\milli\second}$, $g = 1.5$ and $w_l = 1$.
Models were trained for \SI{300}{epochs}, using the PyTorch Lightning package to manage training, resulting in training and validation accuracy of \SI{95}{\percent} and \SI{93}{\percent}.

\begin{figure}
    \centering
    \includegraphics[width=60mm]{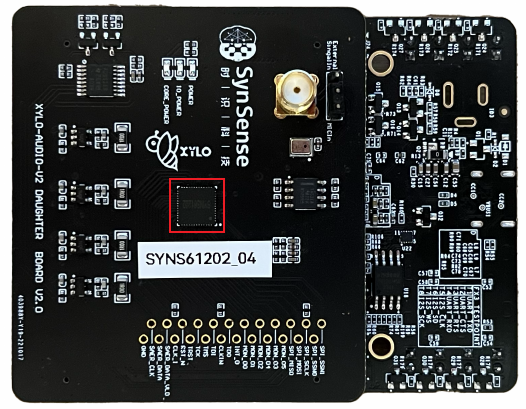}
    \caption{
        \textbf{The Xylo™Audio 2 hardware development kit (HDK).}
        The HDK is a USB bus-power board requiring a PC-host for power and interfacing.
        The HDK interfaces with the open-source Rockpool toolchain for deployment and testing.
        An analog microphone and a analog jack are provided for direct analog single-channel differential input.
        Encoded audio data can alternatively be streamed from the host PC.
        Inference is performed on the Xylo device (red outline).
    }
    \label{fig:xyloa2_hdk}
\end{figure}

\begin{figure}
    \centering
    \includegraphics[width=\linewidth]{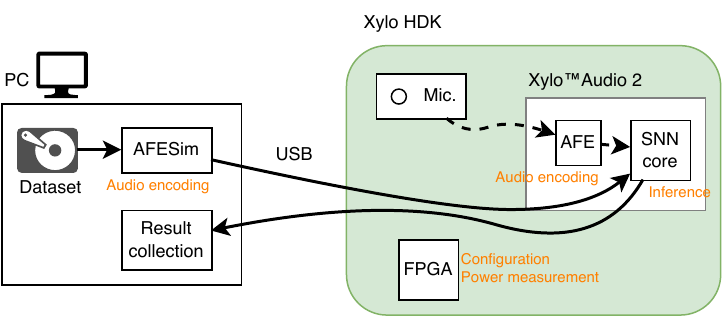}
    \caption{
        \textbf{Benchmarking system overview.}
        The Xylo development kit (``Xylo HDK'', green) is connected via a USB cable to a PC.
        In dataset-driven inference mode, a simulation of the audio encoding block is used to encode audio samples to events (``AFESim'').
        These encoded samples are streamed to the the SNN inference core in the Xylo™Audio 2 device (solid lines).
        Inference is performed entirely on the SNN inference core, and the readout events returned to the PC via usb (solid lines).
        The PC performs an $\textrm{argmax}$ to obtain the inferred class.
        An FPGA on the development kit manages configuration and power measurement.
        In live streaming mode, audio is recorded by a microphone on the development kit, sent to the Audio Front-End (AFE) core in the Xylo™Audio 2 device.
    }
    \label{fig:system_overview}
\end{figure}

\section*{Power, latency and energy per inference}

Trained model was quantized and deployed to Xylo device using the Rockpool deployment pipeline.
Power was measured on the deployed benchmark application, on the XyloAudio 2 hardware development kit (Figure~\ref{fig:xyloa2_hdk}), during streaming continuous analysis of the test-set.
The master clock frequency for XyloAudio was set to \SI{12.5}{\mega\hertz}.
Current measurements were taken using on-board current monitors, at a frequency of \SI{1280}{\hertz}. The consumed power by audio encoder and SNN core were measured in two separate experiments.  

\subsection{Audio encoding power}
The consumed power by audio preprocessing and encoding hardware of XyloAudio device was measured in streaming mode, by playing all test audio samples, resulting in an averaged value (over all \num{16240} test samples of \SI{1}{\second} duration) of \SI {15}{\micro\watt}.
In streaming mode, data is continuously encoded in real-time with no additional latency introduced.

\subsection{Active power and inference latency}
All samples of test dataset were streamed in spike encoded format in accelerated inference mode to Xylo, resulting in test accuracy of \SI{80}{\percent}.
Longer samples (\SI{1}{\second} samples, grouped in sequences of \num{10} samples) were streamed in the same inference mode resulting in averaged inference time of \SI{84}{\milli\second} per sample.
This corresponds to an inference rate of $\approxeq \SI{12}{\hertz}$, and $\SI{12}{\times}$ speedup over real-time.
Power measurements were obtained by current measurement devices integrated in the Xylo development kit, and managed by an on-board FPGA.
Current into the source nets for the XyloAudio device were sampled continuously at \SI{1280}{\hertz}, and converted to power measurements by multiplying by source voltage.
Power measurements were obtained for separate source nets for the SNN inference core and the AFE audio encoding core.

Idle power was measured by deploying the model to the device, then measuring consumed inference core power for five seconds with no model input.
We observed idle power of \SI{351}{\micro\watt}.
The active power of the SNN core was measured by streaming encoded \SI{10}{\second} segments of test dataset to the Xylo device in accelerated inference mode, resulting in continuous average active power consumption of \SI{692}{\micro\watt}.
This results in active energy per inference of \SI{57.6}{\micro\joule\per Inf}.
Dynamic power was estimated as the difference between idle and active power, giving \SI{341}{\micro\watt}.
Dynamic energy per inference is then estimated as \SI{28.4}{\micro\joule\per Inf}.

\begin{table}
    \resizebox{\linewidth}{!}{%
    \begin{tabular}{llllll}
         & Idle & Dyn. & Act. & Dyn. E  & Act. E  \\
         &  (\SI{}{\micro\watt}) & (\SI{}{\micro\watt}) & (\SI{}{\micro\watt}) & (\SI{}{\micro\joule\per Inf}) & (\SI{}{\micro\joule\per Inf}) \\
        \hline
        {Inference} & {351} & {341} & {692} & {28.4} & {57.6} \\
        {Encoding} & & & {15$^\dagger$} & & \\
        \hline \\
    \end{tabular}
    }
    \caption{
        Power and energy measurements for the audio task, measured in accelerated-time mode.
        $^\dagger$Encoding power measured separately to inference, in real-time streaming mode.
    }
    \label{tab:energy-per-inference}
\end{table}


\bibliographystyle{IEEEtran}
\bibliography{bibliography.bib}

\end{document}